\documentclass[11pt]{article}
\usepackage{jheppub}
\usepackage{bbm,bm,mathtools,slashed}
\usepackage[T1]{fontenc}
\usepackage[nottoc]{tocbibind}
\usepackage[toc,page]{appendix}
\usepackage{youngtab}
\usepackage{url}
\usepackage{physics}
\usepackage{comment}
\graphicspath{{./figs/}}

\newcommand{\be}{\begin{equation}}
\newcommand{\ee}{\end{equation}}
\newcommand{\bea}{\begin{eqnarray}}
\newcommand{\eea}{\end{eqnarray}}

\newlength{\fighskip} \fighskip=2pt
\newlength{\figvskip} \figvskip=3pt

\preprint{RIKEN-iTHEMS-Report-24, YITP-24-147}

\title{Phase transition on superfluid vortices in Higgs-Confinement crossover}
\author[a,b]{Tomoya Hayata,}
\author[c,b]{Yoshimasa Hidaka}
\author[d]{and Dan Kondo}

\affiliation[a]{Departments of Physics, Keio University School of Medicine, 4-1-1 Hiyoshi, Kanagawa 223-8521, Japan}
\affiliation[b]{Interdisciplinary Theoretical and Mathematical Sciences Program (iTHEMS), RIKEN, Wako, Saitama 351-0198, Japan}
\affiliation[c]{Yukawa Institute for Theoretical Physics, Kyoto University, Kyoto 606-8502, Japan}
\affiliation[d]{Kavli Institute for the Physics and Mathematics of the Universe (WPI), University of Tokyo Institutes for Advanced Study, University of Tokyo, Kashiwa 277-8583, Japan}

\emailAdd{hayata@keio.jp}
\emailAdd{yoshimasa.hidaka@yukawa.kyoto-u.ac.jp}
\emailAdd{dan.kondo@ipmu.jp}

\abstract{
  We propose a novel method to distinguish states of matter by identifying spontaneous symmetry breaking on extended objects, such as vortices, even in the absence of a bulk phase transition. As a specific example, we investigate the phase transition on superfluid vortices in the Higgs-confinement crossover using a $\mathrm{U}(1)_\mathrm{gauge} \times \mathrm{U}(1)_\mathrm{global}$ model in $(3+1)$ dimensions. This model exhibits superfluidity of $\mathrm{U}(1)_\mathrm{global}$ symmetry and allows for a crossover between the Higgs and confinement regimes by varying the gauge coupling constant from weak to strong. We demonstrate that, on vortices, spontaneous breaking of the $\mathbb{Z}_2$ flavor symmetry occurs in the weak coupling (Higgs) regime, while it does not in the strong coupling (confinement) regime. We also confirm that those regimes are separated by a second-order phase transition through Monte Carlo simulations, whose universality class corresponds to the two-dimensional Ising model.}

\begin{document}
\maketitle

\section{Introduction}
Understanding the phase structure of gauge theories and their physical properties is a fundamental problem in high-energy physics.
In particular, determining the phase structure of QCD at low temperatures and finite densities is one of the most important unsolved problems (See, e.g., \cite{Fukushima:2010bq} for a review).
It is important to understand the physics inside neutron stars, and also, the equation of states of low-temperature and dense QCD is relevant to astrophysics~\cite{Baym:2017whm}.
Since the presence of phase transition strongly affects the stiffness of equation of states, which is important to understand the mass-radius relation of neutron stars, there has been much discussion on a possible phase transition from hadronic to quark matters. According to the classification of phases of matters based on spontaneous symmetry breaking~\cite{Landau:1937obd,Ginzburg:1950sr,Nambu:1960tm,Nambu:1961fr}, the symmetry breaking pattern of three-flavor QCD is the same in both the hadronic superfluid phase and the superconducting phase of quarks (or Higgs phase), known as the color-flavor locked (CFL) phase~\cite{Alford:1997zt,Alford:1998mk,Alford:2007xm}. Therefore, the crossover between them is expected as a function of the baryon chemical potential, which is referred to as quark-hadron continuity~\cite{Schafer:1998ef,Schafer:1999fe}.
In this paper, we refer to this, more generally, as Higgs-confinement continuity or crossover, and do not limit ourselves to finite-density QCD;
such continuity was studied in lattice gauge theory~\cite{Fradkin:1978dv,Banks:1979fi,OSTERWALDER1978440}.
Recently, this problem has been reexamined through the lens of generalized global symmetries and/or topological phases \cite{Cherman:2018jir,Hirono:2018fjr,Hirono:2019oup,Cherman:2020hbe,Hidaka:2022blq,Verresen:2022mcr,Thorngren:2023ple,Chung:2024hsq,Hayashi:2023sas,Cherman:2024exo}, through the matching of vortex excitations~\cite{Alford:2018mqj,Chatterjee:2018nxe,Fujimoto:2019sxg,Fujimoto:2020dsa,Fujimoto:2021wsr,Fujimoto:2021bes}, 
and numerical studies~\cite{Greensite:2017ajx,Greensite:2018mhh,Greensite:2020nhg,Greensite:2021fyi,Ikeda:2023kcf}. 
However, the existence of the phase transition remains an open question.

Our aim in this paper is to investigate the characteristics that distinguish states in the Higgs-confinement crossover.
The excitation spectra and physical properties of confinement and Higgs regimes are different: In the confinement regime, baryons are composite particles with internal excitations, whereas, in the Higgs regime, they behave like point particles of quarks.
However, these excitation states are expected to change continuously without singularity in the Higgs-confinement crossover.
Below, we focus on lattice gauge theories with Higgs fields~\cite{Fradkin:1978dv} in which the confinement and Higgs phases appear in the strong and weak coupling regimes, and they are smoothly connected as parameters of the gauge theory change continuously. Thus, we refer the confinement and Higgs regiemes to the strong and weak coupling regimes, respectively.

In this paper, we propose a classification to distinguish states of matter based on spontaneous symmetry breaking on extended objects, such as topological defects.
This approach could help distinguish states even when there is no bulk phase transition, as the change involves a phase transition on the extended objects themselves.
Examples of extended objects include domain walls and vortices.
These excitations appear in phases where discrete or continuous symmetries are spontaneously broken.
This is indeed the case for three-flavor QCD, where the $\mathrm{U}(1)$ baryon symmetry is spontaneously broken in both the Higgs and confinement regimes, leading to the existence of a superfluid vortex.
Specifically, the CFL vortex is a vortex that exhibits both superconductivity and superfluidity~\cite{Balachandran:2005ev,Nakano:2008dc,Eto:2009kg,Eto:2013hoa,Nakano:2007dr}.

The effective theories of those domain wall and vortex excitations can be described in $d$ and $(d-1)$-dimensions in a $(d+1)$-dimensional theory, respectively.
Consequently, as the bulk parameters change, the parameters of the effective theory also change, which leads to the possibility of a phase transition occurring on the extended objects, even in the absence of a bulk phase transition. Therefore, extended objects such as topological defects can be used as a probe for classifying states of matter like changing the topology of space to detect topological order.

In this paper, we consider a lattice model of a $\mathrm{U}(1)$ gauge theory with two Higgs fields, referred to as the $\mathrm{U}(1)_\mathrm{gauge} \times \mathrm{U}(1)_\mathrm{global}$ model in $(3+1)$ dimensions.
This model exhibits superfluidity, and furthermore, the gauge coupling can be changed from weak to strong coupling.
Without the Higgs fields, this model exhibits confinement and Coulomb phases depending on the strength of the gauge coupling constant.
Through Higgs condensation, the strong and weak coupling regions are expected to be continuously connected in the bulk.
The possibility of a phase transition occurring on vortices in this model was conjectured by Motrunich and Senthil using the (1+1)-dimensional quantum Ising model as an effective theory of low-energy physics of a vortex in ref.~\cite{motrunich2005origin}.
In this paper, we confirm that this phase transition occurs using analytical considerations and ab-initio calculations through Monte Carlo simulations of the $\mathrm{U}(1)_\mathrm{gauge} \times \mathrm{U}(1)_\mathrm{global}$ model.
This result may pave the way for classifying states of matter beyond the Landau-Ginzburg-Wilson theory based on the symmetry-breaking patterns.

The rest of the paper is organized as follows.
In section \ref{sec:U(1)U(1)}, we introduce the $\mathrm{U}(1)_\mathrm{gauge}\times \mathrm{U}(1)_\mathrm{global}$ model and discuss the symmetry breaking on a vortex.
We introduce a symmetry defect to detect spontaneous symmetry breaking, and analytically show that spontaneous breaking of discrete flavor symmetry occurs in the weak coupling regime. 
Section~\ref{sec:U(1)U(1)numericalLattice} presents the details of our lattice numerical simulations based on Monte Carlo methods.
We discuss the numerical results, providing evidence of a second-order phase transition on vortices and its relation to the Ising universality class in two dimensions.
Section~\ref{sec:summary} is devoted to the summary.
Appendix~\ref{sec:Ising} discusses the Ising model as a specific example illustrating symmetry breaking using symmetry defects.


\section{Phase transition on a vortex in \texorpdfstring{$\mathrm{U}(1)_\mathrm{gauge}\times \mathrm{U}(1)_\mathrm{global}$}{U(1)gauge x U(1)global} model}\label{sec:U(1)U(1)}
 
We consider a lattice model with $\mathrm{U}(1)_\mathrm{gauge}\times \mathrm{U}(1)_\mathrm{global}$ symmetry in (3+1) dimensions. 
This model is a simple model that exhibits superfluidity and allows a crossover between the confinement (strong coupling) and Higgs (weak coupling) regimes by varying the gauge coupling constant.
It also possesses a $\mathbb{Z}_{2}$ flavor symmetry that remains unbroken in both the strong and weak coupling regimes. However, we will show that the insertion of a quantum vortex associated with superfluidity induces spontaneous symmetry breaking of the $\mathbb{Z}_{2}$ flavor symmetry on the vortex in the weak coupling regime.
It was first conjectured by Motrunich and Senthil in ref.~\cite{motrunich2005origin} that spontaneous symmetry breaking accompanies the second-order phase transition with the two-dimensional Ising universality class.
In this section, we analytically show that the spontaneous symmetry breaking indeed occurs by showing different patterns of symmetry realization (symmetry breaking patterns) in the strong coupling and weak coupling regimes.
We will confirm its nature as the phase transition numerically by performing Monte Carlo simulations in section~\ref{sec:U(1)U(1)numericalLattice}.

\subsection{\texorpdfstring{$\mathrm{U}(1)_\mathrm{gauge}\times \mathrm{U}(1)_\mathrm{global}$}{U(1)gauge x U(1)global} model}
The action that we consider is 
\begin{equation}\label{eq:action}
S
=-\frac{\beta_H}{2}\sum_{x,\mu}\phi^*_{1}(x)U_\mu(x)\phi_{1}(x+\hat{\mu}) -\frac{\beta_H}{2}\sum_{x,\mu}\phi^*_{2}(x)U_\mu(x)\phi_{2}(x+\hat{\mu})-\frac{\beta_g}{2}\sum_{x,\mu<\nu}U_{\mu\nu}(x) + \; \text{c.c.} , 
\end{equation}
or equivalently, it can be expressed as
 \begin{equation}\label{eq:Action_A}
 S
 =-\beta_H\sum_{x,\mu}\qty[\cos(\Delta_\mu\theta_{1}(x)+A_\mu(x))+\cos(\Delta_\mu\theta_{2}(x)+A_\mu(x)) ]-\beta_g\sum_{x,\mu<\nu}\cos(F_{\mu\nu}(x)).
\end{equation}
Here, $\phi_1(x)=\exp[i\theta_1(x)]$ and $\phi_2(x)=\exp[i\theta_2(x)]$ are $\mathrm{U}(1)$-valued charged scalar fields that represent the phase component of the Higgs fields, and $\beta_H$ is the hopping parameter.
This setup corresponds to assuming a deep Higgs potential, where the radial parts of the Higgs fields are fixed, and with magnitude $|\phi_a|$ absorbed into the strength of $\beta_H$.
$U_\mu(x)=\exp[iA_\mu(x)]$ is the $\mathrm{U}(1)$ link variable with gauge field $A_\mu(x)$, defined on a link between sites $x$ and $x+\hat{\mu}$, where $\hat{\mu}$ is the unit vector in the $\mu$ direction.
$\beta_g$ is the coupling constant that is inversely proportional to the square of gauge coupling.
Therefore, we refer to small and large values of $\beta_g$ as strong and weak coupling, respectively.
The discrete derivative is given by $\Delta_\mu\theta_{a}(x)\coloneqq\theta_{a}(x+\hat{\mu})-\theta_{a}(x)$,
and the plaquette $U_{\mu\nu}(x)$ is defined as
\begin{equation}
    U_{\mu\nu}(x)\coloneqq U_\mu(x)U_\nu(x+\hat{\mu})U_\mu^\dag(x+\hat{\nu})U_\nu^\dag(x)=\exp(iF_{\mu\nu}),
\end{equation}
where $F_{\mu\nu}\coloneqq\Delta_\mu A_\nu-\Delta_\nu A_\mu$.
We label four-dimensional spacetime discretized on a hypercubic lattice $x$ as $x=(x_1,x_2,x_3,x_4)$, and $x_\mu$ run $0$, $1$, $\ldots N_\mu-1$. The lattice volume is given by $V=N_1N_2N_3N_4$.
For later convenience, we parametrize $x$ as $x=(x_\perp,z)$ with $x_\perp=(\tau, x_1,x_2)$, $z=x_3$ and $\tau=x_4$,
and rewrite the action using the sum-to-product formula as 
\begin{equation}
  \begin{split}
    S
    &=-2\beta_H\sum_{x,\mu}
    \cos(\frac{\Delta_\mu\theta_{1}(x)-\Delta_\mu\theta_{1}(x)}{2})\cos(\frac{\Delta_\mu\theta_{1}(x)+\Delta_\mu\theta_{2}(x)}{2}+A_\mu(x))\\
    &\quad-\beta_g\sum_{x,\mu<\nu}\cos(F_{\mu\nu}(x))
    .
  \end{split}\label{eq:Action_product}
\end{equation}

This model has $\mathrm{U}(1)_{\mathrm{gauge}}\times\mathrm{U}(1)_{\mathrm{global}}$ symmetry\footnote{Note that we assign the opposite charge to $\phi_2$ compared to refs.~\cite{Cherman:2020hbe, Hidaka:2022blq, Hayashi:2023sas,Cherman:2024exo}, where the charges of $\phi_2$ under $\mathrm{U}(1)_\mathrm{gauge}$ and $\mathrm{U}(1)_\mathrm{global}$ are $-1$ and $+1$, respectively.}
\begin{align}
\mathrm{U}(1)_{\mathrm{gauge}}&: (\phi_{1},\phi_{2})\rightarrow (\mathrm{e}^{i\lambda}\phi_1,\mathrm{e}^{i\lambda}\phi_2), \\   
\mathrm{U}(1)_{\mathrm{global}}&: (\phi_{1},\phi_{2})\rightarrow (\mathrm{e}^{i\eta}\phi_1,\mathrm{e}^{-i\eta}\phi_2),
\end{align}
where $\lambda$ is the gauge parameter depending on spacetime coordinate, and $\eta$ is constant.
Since the transformations $\lambda=\pi$ and $\eta=\pi$ leave $\phi_1$ and $\phi_2$ invariant, this transformation is redundant.
Consequently, the global symmetry is not $\mathrm{U}(1)_{\mathrm{global}}$ but $\mathrm{U}(1)_{\mathrm{global}}/\mathbb{Z}_2$.
In addition to the continuous symmetry, there are two types of discrete symmetries:
\begin{align}
    \text{$\mathbb{Z}_2$ flavor symmetry}&: \phi_1 \rightarrow \phi_2, \quad \phi_2 \rightarrow \phi_1, \quad  U_\mu \rightarrow U_\mu,\label{eq:flavor_symmetry}\\
    \text{Charge conjugation}&: \phi_i \rightarrow \phi^*_i, \quad U_\mu \rightarrow U^*_\mu.
\end{align}
The $\mathbb{Z}_2$ flavor symmetry will play an important role in our study of phase transition on a vortex.

We define the expectation value of an operator $\mathcal{O}$, using the path integral as
\begin{align}
    \expval{\mathcal{O}}&=\frac{1}{Z}\int DA_\mu D\theta_1 D\theta_2 \mathrm{e}^{-S}\mathcal{O},
\end{align}
where $Z$ is the partition function defined as 
\begin{equation}
    Z\coloneqq\int DA_\mu D\theta_1 D\theta_2 \mathrm{e}^{-S}.
\end{equation}
Here, $DA_\mu=\prod_x dA_\mu(x)/2\pi$, $D\theta_1=\prod_x d\theta_1(x)/2\pi$ and $D\theta_2=\prod_x d\theta_2(x)/2\pi$ are the Haar measures for the $\mathrm{U}(1)$ gauge field and $\mathrm{U}(1)$ valued scalar fields, respectively. These measures are normalized such that $\int DU_{\mu}=\int D\phi_1=\int D\phi_2=1$.

Note that our model slightly differs from that in previous studies on confinement-Higgs continuity or transition, as discussed in refs.~\cite{Cherman:2020hbe, Hidaka:2022blq, Hayashi:2023sas,Cherman:2024exo}. These studies introduce a neutral field that mimics a baryon and also consider a superfluid phase induced by the condensation of this neutral field.

\subsection{\texorpdfstring{$\mathbb{Z}_2$}{Z2} flavor symmetry defect and symmetry breaking}
In this subsection, we introduce a criterion using a symmetry defect to detect the spontaneous breaking of discrete symmetry.
An ordinary symmetry operator is a topological object defined on a constant time slice.
When the topological object is extended in the time direction, it is called a symmetry defect, which corresponds to imposing twisted boundary conditions.

Spontaneous symmetry breaking is characterized by a non-vanishing order parameter or by the long-distance limit of the two-point function of the order operator~\cite{Weinberg:1996kr}.
Alternatively, in the case of discrete symmetries, spontaneous breaking can be characterized by the behavior of the partition function in the presence of a symmetry defect (disordered parameter), where
\begin{equation}\label{eq:criteria}
\lim_{V\to\infty}\frac{Z'}{Z}\to 
\begin{cases}
 \sim e^{-N_4 V_\parallel T_w}\to 0& \text{ broken phase}\\
 \text{const}\neq 0 &  \text{unbroken phase}
\end{cases}.
\end{equation}
Here, $Z$ and $Z'$ are partition functions without and with a symmetry defect, respectively.

$V_\parallel$ is the volume parallel to the domain wall, and $T_w$ is the domain wall tension.
In systems where the discrete symmetry is spontaneously broken, the insertion of the symmetry defect induces the domain wall connecting the different vacua. The energy of the domain wall leads to an exponential suppression in the ratio of the partition function to that in the absence of the domain wall.
In contrast, in phases without symmetry breaking, the partition function remains insensitive to the boundary conditions. 
As a result, the ratio of the partition functions is given by  eq.~\eqref{eq:criteria}.
In Appendix \ref{sec:Ising}, we present the quantum Ising model as an example that can be solved analytically.

We insert a $\mathbb{Z}_2$ flavor symmetry~\eqref{eq:flavor_symmetry} defect on a plane perpendicular to the $z$-axis. 
Specifically, we introduce the defect between $z=N_3-1$ and $z=N_3$ $(=0)$.

The insertion of this defect modifies the matter field term in the action at $z=N_3-1$ and $\mu=z$ as 
\begin{equation}
  \begin{split}
    &\cos(\Delta_z\theta_{1}(x_\perp,N_3-1)+A_z(x_\perp,N_3-1))+\cos(\Delta_z\theta_{2}(x_\perp,N_3-1)+A_z(x_\perp,N_3-1))\\
    &\quad\to
    \cos(\theta_{2}(x_\perp,0)-\theta_{1}(x_\perp,N_3-1)+A_z(x_\perp,N_3-1))\\
    &\qquad+\cos(\theta_{1}(x_\perp,0)-\theta_{2}(x_\perp,N_3-1)+A_z(x_\perp,N_3-1)).
  \end{split}
\end{equation}
The other terms remain unchanged. As a result, the action with the symmetry defect $S'$ is given by
\begin{equation}\label{eq:S'}
  \begin{split}
    S' &=
    -\beta_H\sum_{x_\perp}\Bigl[\cos(\theta_{2}(x_\perp,0)-\theta_{1}(x_\perp,N_3-1)+A_z(x_\perp,N_3-1))\\
    &\qquad\qquad\quad+\cos(\theta_{1}(x_\perp,0)-\theta_{2}(x_\perp,N_3-1)+A_z(x_\perp,N_3-1))\\
    &\qquad\qquad\quad+\sum_{z\neq N_3-1}\qty[\cos(\Delta_z\theta_{1}(x)+A_z(x))+\cos(\Delta_z\theta_{2}(x)+A_z(x)) ]\\
    &\qquad\qquad\quad+\sum_{z,\mu\neq z}\qty[\cos(\Delta_\mu\theta_{1}(x)+A_\mu(x))+\cos(\Delta_\mu\theta_{2}(x)+A_\mu(x)) ]\Bigr]
    -\beta_g\sum_{x,\mu<\nu}\cos(F_{\mu\nu}(x))\\
    = &S
    -\beta_H\sum_{x_\perp}\Bigl[\cos(\theta_{2}(x_\perp,0)-\theta_{1}(x_\perp,N_3-1)+A_z(x_\perp,N_3-1))\\
    &\qquad\qquad\quad+\cos(\theta_{1}(x_\perp,0)-\theta_{2}(x_\perp,N_3-1)+A_z(x_\perp,N_3-1))\\
      &\qquad\qquad\quad -
      \cos(\Delta_z\theta_{1}(x_\perp,N_3-1)+A_z(x_\perp,N_3-1))\\
    &\qquad\qquad\quad-\cos(\Delta_z\theta_{2}(x_\perp,N_3-1)+A_z(x_\perp,N_3-1))
      \Bigr].
  \end{split}
\end{equation}

\subsection{Bulk phase structure}
Let us first consider the case without vortices, where we will show that the $\mathbb{Z}_2$ flavor symmetry remains unbroken in both the strong and weak coupling regimes.
\subsubsection{Strong coupling limit}
In the strong coupling limit $\beta_g=0$, we can directly integrate out the gauge field $A_\mu$ in the partition function,

\begin{align}
    \label{eq:Z}
    Z&=\int DA_\mu D\theta_1 D\theta_2 \exp\Biggl[\notag\\
    &\qquad 2\beta_H \sum_{x,\mu}   \cos(\frac{\Delta_\mu\theta_{1}(x)-\Delta_\mu\theta_{2}(x)}{2}) \cos( \frac{\Delta_\mu\theta_{1}(x)+\Delta_\mu\theta_{2}(x)}{2} +A_\mu(x)) \Biggr]\nonumber\\
    &=\int D\theta_1 D\theta_2 \prod_{x,\mu}I_0\left[2\beta_H   \cos(\frac{\Delta_\mu\theta_{1}(x)-\Delta_\mu\theta_{2}(x)}{2})  \right],
\end{align}
where $I_0(z)$ is the modified Bessel function.
Since the partition function depends only on $\theta_1-\theta_2$, we can eliminate $\theta_2$ by redefining the field  $\theta'_1= \theta_1-\theta_2$.
Then, the partition function becomes
\begin{equation}\label{eq:Z_strong}
    Z=\int D\theta'_1 \prod_{x,\mu}I_0\left[2\beta_H   \cos(\frac{\Delta_\mu\theta'_1(x)}{2})  \right].
\end{equation}
We are interested in the region where $\beta_H$ is large, and evaluate the partition function using the saddle-point approximation.
Since $I_0(z)$ is both even and convex, we aim to maximize $|\cos({\Delta_\mu\theta'_1(x)}/2)|$.
This is achieved when $\Delta_\mu \theta'_1 = 0 \mod 2\pi$.
Considering that $\theta'_1$ and $\theta'_1 + 2\pi$ represent the same point, it is sufficient to consider only $\Delta_\mu \theta'_1 = 0$. Thus, the solution is that $\theta'_1(x)$ is constant, and the partition function becomes 
\begin{equation}
    Z\approx \qty[I_0(2\beta_H)]^{4V},
\end{equation}
where ``$\approx$'' represents the saddle-point approximation, and $V$ represents the lattice volume. The factor $4$ comes from the four directions of $\mu$.

When we insert a $\mathbb{Z}_{2}$ symmetry defect, we obtain the partition function by integrating out the gauge field,
\begin{equation}
  \begin{split}
    \label{eq:Z_symmetrydefect}
    Z' &=\int D\theta_1 D\theta_2 \prod_{x,\mu\neq z}I_0\qty[2\beta_H \cos(\frac{\Delta_\mu\theta_{1}(x)-\Delta_\mu\theta_{2}(x)}{2}  ) ]
    \\
    &\qquad\times    \prod_{x_\perp,z\neq N_3-1}I_0\qty[2\beta_H \cos(\frac{\Delta_z\theta_{1}(x)-\Delta_z\theta_{2}(x)}{2}  ) ]
     \\
    &\qquad\times\prod_{x_\perp}I_0\qty[2\beta_H   \cos(\frac{
        \theta_{2}(x_\perp,0)-\theta_{1}(x_\perp,N_3-1)
        -\qty(\theta_{1}(x_\perp,0)-\theta_{2}(x_\perp,N_3-1))}{2}  )  ].
  \end{split}
\end{equation}
Again, we define  $\theta'_1(x)\coloneqq \theta_1(x)-\theta_2(x)$, and obtain
\begin{equation} 
    \begin{split}\label{eq:Z'_strong}
      Z' &=\int D\theta'_1 D\theta_2 \prod_{x,\mu\neq z}I_0\qty[2\beta_H \cos(\frac{\Delta_\mu\theta'     _{1}(x)}{2})]
      \prod_{x_\perp,z\neq N_3-1}I_0\qty[2\beta_H   \cos(\frac{\Delta_z\theta'_{1}(x)}{2})  ]
       \\
      &\qquad\times\prod_{x_\perp}I_0\qty[2\beta_H   \cos(\frac{
        2\theta_{2}(x_\perp,0)+2\theta_{2}(x_\perp,N_3-1)-\theta'_{1}(x_\perp,0)-\theta'_{1}(x_\perp,N_3-1)}{2}  )  ]\\
        &=\int D\theta'_1 D\theta'_2 \prod_{x,\mu\neq z}I_0\qty[2\beta_H \cos(\frac{\Delta_\mu\theta'_{1}(x)}{2})]
        \prod_{x_\perp,z\neq N_3-1}I_0\qty[2\beta_H   \cos(\frac{\Delta_z\theta'_{1}(x)}{2})  ]
         \\
        &\qquad\times\prod_{x_\perp}I_0\qty[2\beta_H   \cos(
          \theta'_{2}(x_\perp,N_3-1)  )  ].
    \end{split}
  \end{equation}
In the second line, we defined $\theta'_2(x_\perp,N_3-1)\coloneqq \theta_2(x_\perp,0)+\theta_2(x_\perp,N_3-1)-\theta'_1(x_\perp,0)/2-\theta'_1(x_\perp,N_3-1)/2$.
The classical solution is $\theta'_1=\text{const}$ and $\theta'_2(x_\perp,N_3-1)=0,\pi$,
so the partition function becomes
\begin{equation}
    Z'\approx \qty[I_0(2\beta_H)]^{4V}.
\end{equation}
Within the saddle-point approximation, the ratio of the partition function, with and without symmetry defect, is given by
\begin{equation}
    \frac{Z'}{Z} \approx 1.
\end{equation}
Therefore, the $\mathbb{Z}_2$ flavor symmetry is unbroken in the strong coupling limit.

\subsubsection{Weak coupling}
In the case of weak coupling $\beta_g\gg1$, we would first like to minimize the plaquette term in the action.
This can be achieved by choosing the gauge field configuration as pure gauge $A_\mu(x)=\Delta_\mu\alpha(x)$, where $\alpha$ is a real-valued function.
Then, the action~\eqref{eq:Action_A} becomes
\begin{equation}
  \begin{split}
    S
    &=-\beta_g P-\beta_H \sum_{x,\mu} \qty[\cos(\Delta_\mu\theta_{1}+\Delta_\mu\alpha)+\cos(\Delta_\mu\theta_2+\Delta_\mu\alpha)]\\
    &=-\beta_g P-\beta_H \sum_{x,\mu} \qty[\cos(\Delta_\mu\theta_{1})+\cos(\Delta_\mu\theta_2)],
  \end{split}
\end{equation}
where $P$ is the number of Plaquettes.
In the second line, we shifted $\theta_1(x)\to \theta_1(x)-\alpha(x)$ and $\theta_2(x)\to \theta_2(x)-\alpha(x)$ to remove the gauge field.
Minimizing the matter part of the action results in $\theta_1(x)=\text{const}$ and $\theta_2(x)=\text{const}$;
then, the partition function becomes
\begin{align}
    Z&\approx \exp(\beta_g P+ 8V\beta_H).
\end{align} 

When we insert the $\mathbb{Z}_{2}$ flavor symmetry defect,
The action for $\beta_g\gg 1$ becomes
  \begin{equation}
    \begin{split}
      S' &=
      -\beta_g P
      -\beta_H\sum_{x_\perp}\Bigl[\cos(\theta_{2}(x_\perp,0)-\theta_{1}(x_\perp,N_3-1))+\cos(\theta_{1}(x_\perp,0)-\theta_{2}(x_\perp,N_3-1))\\
      &\quad+\sum_{z\neq N_3-1}\qty[\cos(\Delta_z\theta_{1}(x))+\cos(\Delta_z\theta_{2}(x)) ]
      +\sum_{z,\mu\neq z}\qty[\cos(\Delta_\mu\theta_{1}(x))+\cos(\Delta_\mu\theta_{2}(x)) ]\Bigr].
    \end{split}
  \end{equation}
Here, we used the same approximation for the gauge field as in the case without the symmetry defect.
Minimizing the matter part gives
$\Delta_\mu\theta_1(x)=0$, $\Delta_\mu\theta_2(x)=0$, $\theta_2(x_\perp,0)=\theta_1(x_\perp,N_3-1)$, and $\theta_1(x_\perp,0)=\theta_2(x_\perp,N_3-1)$.
The solution is $\theta_1(x)=\theta_2(x)=\text{const}$, which leads to
\begin{equation}
    Z'\approx  \exp(\beta_g P+ 8V\beta_H).
\end{equation}
Therefore, the ratio of the partition functions becomes
\begin{equation}
    \frac{Z'}{Z}\approx 1,
\end{equation}
which implies that the $\mathbb{Z}_2$ symmetry is unbroken in the weak coupling regime.

We showed that the $\mathbb{Z}_2$ flavor symmetry remains unbroken in both the strong and weak coupling regimes.
While this does not necessarily imply the absence of a phase transition, it is consistent with continuity.

In the next subsection, we will demonstrate that inserting a vortex probes spontaneous symmetry breaking on the vortex, by showing different symmetry realizations in the strong and weak coupling regimes.

\subsection{Phase transition on a vortex}
Since we are considering the superfluid phase with a large $\beta_H$, quantum vortices may exist in this phase.
We define the winding number for $\theta_a$ ($a=1,2$) as
\begin{equation}\label{eq:circulation_lattice}
  w_a = \int_C \frac{d\theta_a}{2\pi}\coloneqq\sum_{\{x,\mu\}\in{C}}\frac{\Delta_\mu \theta_a}{2\pi} ,
\end{equation}
where $C$ is a closed path.
However, this quantity is not gauge-invariant because
it transforms as $d\theta_a\to d\theta_a+d\lambda$, which shifts $w_a\to w_a+n$, where $n=\int_C d\lambda/2\pi$ is an integer.
The gauge-invariant combinations~\cite{Hidaka:2022blq} are
\begin{equation}
  Q(C)\coloneqq\frac{1}{2\pi}\int_C(d\theta_1-d\theta_2)=w_1-w_2,
\end{equation}
and
\begin{equation}
  W(C)=\exp[i \frac{1}{2}\int_C(d\theta_1+d\theta_2)]=\exp[i\pi(w_1+w_2)].
\end{equation}
These are the charge of $\mathrm{U}(1)$ $2$-form global symmetry, which represents the winding number of vortices, and the symmetry operator of $\mathbb{Z}_2$ $2$-form global symmetry, respectively.

$W(C)$ is related to Wilson loop $\exp(-i \int_C A)$
through the conditions $\Delta_\mu\theta_1+A_\mu=0$ and $\Delta_\mu\theta_2+A_\mu=0$ that minimize the action for large $\beta_H$. The Wilson loop measures the $\mathbb{Z}_2$-valued Aharonov–Bohm phase.
Note that the emergent $\mathbb{Z}_2$ $2$-form global symmetry is not spontaneously broken, so that there exists no topologically ordered phase~\cite{Hirono:2018fjr,Hidaka:2022blq}.

In this subsection, we consider a static $\mathrm{U}(1)_{\mathrm{global}}$ vortex extended along the $z$-axis and positioned at $(x_1,x_2)=(1/2,1/2)$ with the winding number $1$, i.e.,  
\begin{equation}
Q(C)= w_1-w_2=1,
\end{equation}
for all closed paths $C$ encircling the core of a vortex located at $(x_1,x_2)=({1}/{2},{1}/{2})$.
We show that the $\mathbb{Z}_2$ flavor symmetry is spontaneously broken in the weak coupling regime, while it is unbroken in the strong coupling limit.

Note that for $Q(C)=1$, we have $W(C)=-1$ for all values of the coupling constant in this model. Thus, $W(C)$ or the Wilson loop itself cannot capture the phase transition of the bulk or the vortex between strong and weak coupling regimes.

 \subsubsection{Strong coupling limit \texorpdfstring{$\beta_g= 0$}{betag=0}}

The winding number $Q(C)$ depends only on $\theta'_1 = \theta_1 - \theta_2 $, which allows us to start from the partition function in eq.~\eqref{eq:Z_strong}.
Let $\Theta(x_1,x_2)$ be a vortex solution with $Q(C)=1$ that maximizes the following functional in two dimensions:
\begin{equation}
\prod_{x_1,x_2,\mu=1,2}I_0\left[2\beta_H   \cos(\frac{\Delta_\mu\Theta(x_1,x_2)}{2})  \right].
\end{equation}
The specific functional form of the solution is not necessary.
It is sufficient to assume its existence here.
The solution $\Theta(x_1,x_2)$ can be used as the solution of the saddle-point approximation for the partition function as $\theta_1'(x)=\Theta(x_1,x_2)$,
where the partition function in the absence of the symmetry defect~\eqref{eq:Z_strong} becomes
\begin{align}
  Z&\approx\prod_{x}I_0\qty[2\beta_H   \cos(\frac{\Delta_1\Theta(x_1,x_2)}{2})]
  I_0\qty[2\beta_H   \cos(\frac{\Delta_2\Theta(x_1,x_2)}{2}) ]
  I_0\qty(2\beta_H)^{2}.
\end{align}
Here, we used $\Delta_z\Theta(x_1,x_2)=\Delta_\tau\Theta(x_1,x_2)=0$,
which results in the last term, $I_0\qty(2\beta_H)^{2}$.

In the presence of the symmetry defect, the classical solution becomes $\theta_1'(x)=\Theta(x_1,x_2)$, and $\theta'_2(x)=0,\pi$, and the partition function~\eqref{eq:Z'_strong} is
\begin{equation} 
  \begin{split}
    Z'
      &\approx\Biggl[\prod_{x}
      I_0\qty[2\beta_H \cos(\frac{\Delta_1\Theta_{1}(x_1,x_2)}{2})]
      I_0\qty[2\beta_H \cos(\frac{\Delta_2\Theta_{1}(x_1,x_2)}{2})]
      I_0\qty(2\beta_H)\Biggl]
      \\
      &\quad\times\Biggl[\prod_{x_\perp,z\neq N_3-1}I_0\qty(2\beta_H)\Biggr]\Biggl[\prod_{x_\perp}I_0\qty(2\beta_H)\Biggr]\\
      &=\prod_{x}I_0\qty[2\beta_H   \cos(\frac{\Delta_1\Theta(x_1,x_2)}{2})]
      I_0\qty[2\beta_H   \cos(\frac{\Delta_2\Theta(x_1,x_2)}{2})]
      I_0\qty(2\beta_H)^{2},
  \end{split}
\end{equation}
and thus, we obtain
\begin{equation}
    \frac{Z'}{Z}\approx 1,
\end{equation}
which implies that $\mathbb{Z}_2$ flavor symmetry is unbroken in the strong coupling limit.

  Here, the degree of freedom is taken as $\theta'_1$.
  In terms of the degrees of freedom of $\theta_1$ and $\theta_2$, $Q(C)=1$ implies vortices with $w_1=n$  and  $w_2=1-n$ also have the same winding number $Q(C)$. These configurations can be considered as contributing equally to the partition function.

\subsubsection{Weak coupling \texorpdfstring{$\beta_g\gg1$}{betag>>1}}
In the weakly coupled Higgs regime, under the continuous field approximation, vortex solutions exist where the vortex can wind around either $\theta_1$ or $\theta_2$~\cite{Cherman:2020hbe,Cherman:2024exo}.
These vortices carry magnetic flux and exhibit nontrivial gauge configurations.
Solutions with higher winding numbers, where $|w_1|+|w_2|>1$ with fixed $Q(C)=1$, also exist; however, these solutions have higher energy.
Therefore, it is sufficient to consider solutions with $|w_1|=1$ or $|w_2|=1$.
Since the winding number is fixed at $Q(C)=1$, the possible winding number of each vortex is $w_1=1, w_2=0$ or $w_1=0, w_2=-1$.
Let $\theta_1(x)=\tilde{\Theta}(x_1,x_2)$, $\theta_2(x)=0$, $A_\tau(x)=A_z(x)=0$, $A_1(x)=h_1(x_1,x_2)$, $A_2=h_2(x_1,x_2)$
be the solution that minimizes the action~\eqref{eq:Action_A} with $Q(C)=1$.
From this solution, we can construct another solution with $\theta_1(x)=0$, $\theta_2(x)=-\tilde{\Theta}(x_1,x_2)$, $A_\tau(x)=A_z(x)=0$, $A_1(x)=-h_1(x_1,x_2)$, $A_2=-h_2(x_1,x_2)$.
We denote the action of the solution as $S_\mathrm{cl}$.
Then, the partition function is given by
\begin{equation}
  Z\approx 2e^{-S_\mathrm{cl}}.
\end{equation}

Next, let us consider the case where a symmetry defect is inserted. 
Since $(x_1,x_2)$-plane remains unchanged in the presence of the defect, the vortex solution $\tilde{\Theta}(x_1,x_2)$ also stays the same.
However, along the $z$-axis, to minimize the action, it becomes necessary to flip the vortex configuration at a certain point.
This situation is analogous to a domain wall in the Ising model (see Appendix~\ref{sec:Ising} for details). Specifically, we have
\begin{align}
 \theta_1(x) &= \tilde{\Theta}(x_1,x_2)H(z-k),\\
 \theta_2(x) &= -\tilde{\Theta}(x_1,x_2)(1-H(z-k)),
\end{align}
and 
\begin{align}
  \theta_1(x) &= \tilde{\Theta}(x_1,x_2)(1-H(z-k)),\\
  \theta_2(x) &= -\tilde{\Theta}(x_1,x_2)H(z-k),
 \end{align}
where $H(z)$ is the step function
\begin{equation}
  H(z)=
\begin{cases}
  0 & z<0\\
  1 & z\geq0
\end{cases}.
\end{equation}
Inserting these configurations into the action in the presence of symmetry defect~\eqref{eq:S'}, the classical value of the action $S'_\mathrm{cl}$ becomes
\begin{equation}
  S'_\mathrm{cl} = S_\mathrm{cl}
  +2\beta_HN_4 \sum_{x_1,x_2}\Bigl[1-\cos(\Theta(x_1,x_2))] 
    \Bigr].
\end{equation}
Therefore, we find 
\begin{equation}
\frac{Z'}{Z}\approx \frac{2N_3e^{-S'_\mathrm{cl}}}{2e^{-S_\mathrm{cl}}}=N_3 e^{-2\beta_H N_4 \sum_{x_1,x_2}\bigl[1-\cos(\Theta(x_1,x_2)) \bigr]}\to 0,
\end{equation}
which implies that the $\mathbb{Z}_2$ flavor symmetry is spontaneously broken in the weak coupling regime.
In the strong coupling and weak coupling regimes, the pattern of spontaneous symmetry breaking differs. 
Motrunich and Senthil argued that there exists the second-order phase transition in between them by constructing an effective Ising model defined on a vortex ~\cite{motrunich2005origin}.
This implies that a phase transition occurs on the vortex and the superfluid phases are distinguishable even though there is no phase transition in the bulk.

\section{Lattice simulation}\label{sec:U(1)U(1)numericalLattice}
\begin{figure}[t]
  \centering\includegraphics[width=0.6\linewidth]{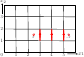}
  \caption{Configuration of the external gauge field $a_\mu$ on the $(x_1,x_2)$ plane for $N_1=6$ and $N_2=4$ shown as an example.
  The red lines indicate the links where a nonvanishing $a_2$ is present. The vortex and anti-vortex are located at $((N_1-1)/2,(N_2-1)/2)=(5/2,1/2)$ and $(N_1-1/2,(N_2-1)/2)=(11/2,1/2)$, respectively.}
  \label{fig:voretex_configuration}
\end{figure}
We perform a lattice simulation to nonperturbatively study the aforementioned phase transition that occurs only on a vortex, not in the bulk state.
The action we consider is given as
\begin{equation}
  \begin{split}
S&=
-\beta_H \sum_{x,\mu}\qty[\cos(\Delta_\mu\theta_1(x)+A_\mu(x)- a_\mu(x))
+\cos(\Delta_\mu\theta_2(x)+A_\mu(x)+ a_\mu(x))]\\
&\quad-\beta_g \sum_{x,\mu<\nu}\cos(F_{\mu\nu}(x)) ,
  \end{split}
\end{equation}
where external fields $a_\mu(x)$ are applied to the original action~\eqref{eq:Action_A} to simulate the ground state in the presence of vortices.
We employ periodic boundary conditions, and set
\be
a_\mu(x)=
\begin{cases}
        \pi \;\; \text{if}\;\; \mu=2, x_1>\frac{N_1}{2}-1, x_2=\frac{N_2}{2}-1 \\
        0 \;\; \text{otherwise} 
\end{cases} 
\ee
that induces a nonzero flux $f_{12}(x)=a_1(x)+a_2(x+\hat{1})-a_1(x+\hat{2})-a_2(x)=\pi$ at $(x_1,x_2)=({N_1}/{2}-1,{N_2}/{2}-1)$, and $f_{12}(x)=-\pi$ at $(x_1,x_2)=(N_1-1,{N_2}/{2}-1)$ with arbitrary $x_3$ and $x_4$, so that world sheet of vortex and that of anti-vortex are fixed at $(x_1,x_2)=((N_1-1)/{2},(N_2-1)/{2})$ and $(x_1,x_2)=(N_1-{1}/{2},{(N_2-1)}/{2})$, respectively.
Figure~\ref{fig:voretex_configuration} shows an example of the configuration of $a_\mu(x)$ and the positions of the vortices for $N_1=6$ and $N_2=4$.

\begin{figure}[t]
  \centering
  \includegraphics[width=0.7\linewidth]{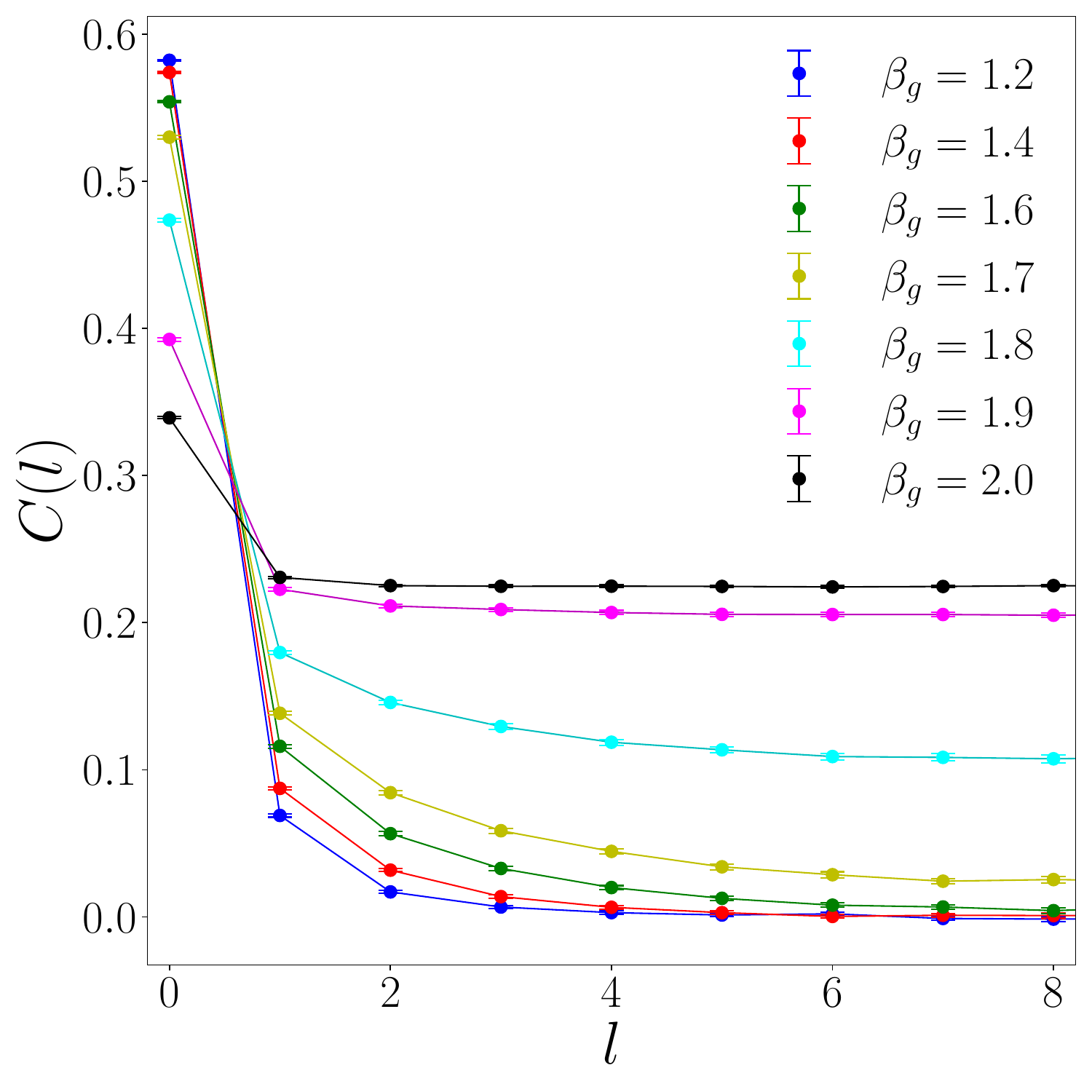}
  \caption{Correlation function of the magnetic flux $\Gamma$. The correlation function approaches nonzero values at a long distance above a critical coupling constant, indicating the long-range order that is observed only in the presence of a vortex.
  }
  \label{fig:LRO}
\end{figure}

Using Monte Carlo simulations, we computed the correlation function of the logarithm of  Wilson loops encircling a vortex to probe the fluctuations of domain walls.
We consider a minimal Wilson loop encircling a vortex placed at $(x_1,x_2)=({(N_1-1)}/{2},{(N_2-1)}/{2})$, 
\begin{equation}
    W(x_3,x_4)=e^{i\oint A}=e^{iF_{12}(\frac{N_1}{2}-1,\frac{N_2}{2}-1,x_3,x_4)},
\end{equation}
and define a magnetic flux $\Gamma(x_3,x_4)$ as 
\be
\Gamma(x_3,x_4)=\frac{1}{i\pi}\log W(x_3,x_4) ,
\ee
and its correlation function $C(l)$ as
\be
C(l)=\frac{1}{N_3N_4}\sum_{x_3,x_4}\langle\Gamma(x_3+l,x_4)\Gamma(x_3,x_4)\rangle.
\ee

Similarly, we can define the magnetic flux and its correlation function for the anti-vortex located at $(x_1,x_2)=(N_1-{1}/{2},{(N_2-1)}/{2})$.
We compute the correlation function both for vortex and anti-vortex, and show the average of the results in figure~\ref{fig:LRO}.
The lattice volume is $V=16^4$. We take $\beta_H=2.0$ and change $\beta_g$ from strong to weak coupling. Under these parameters, we do not observe the phase transition between the confinement and Higgs regime, i.e., no phase transition is detected in the lens of bulk observables.
As clearly seen in figure~\ref{fig:LRO}, we observe the long-range order of $\Gamma(x_3,x_4)$, i.e., the correlation approaches nonzero values at long distance above a critical coupling constant.
Such a long-range order is measurable only in the presence of a vortex, which is in stark contrast to the ordinary phases of matters characterized by spontaneous symmetry breaking.

\begin{figure}[t]
  \centering
  \includegraphics[width=0.7\linewidth]{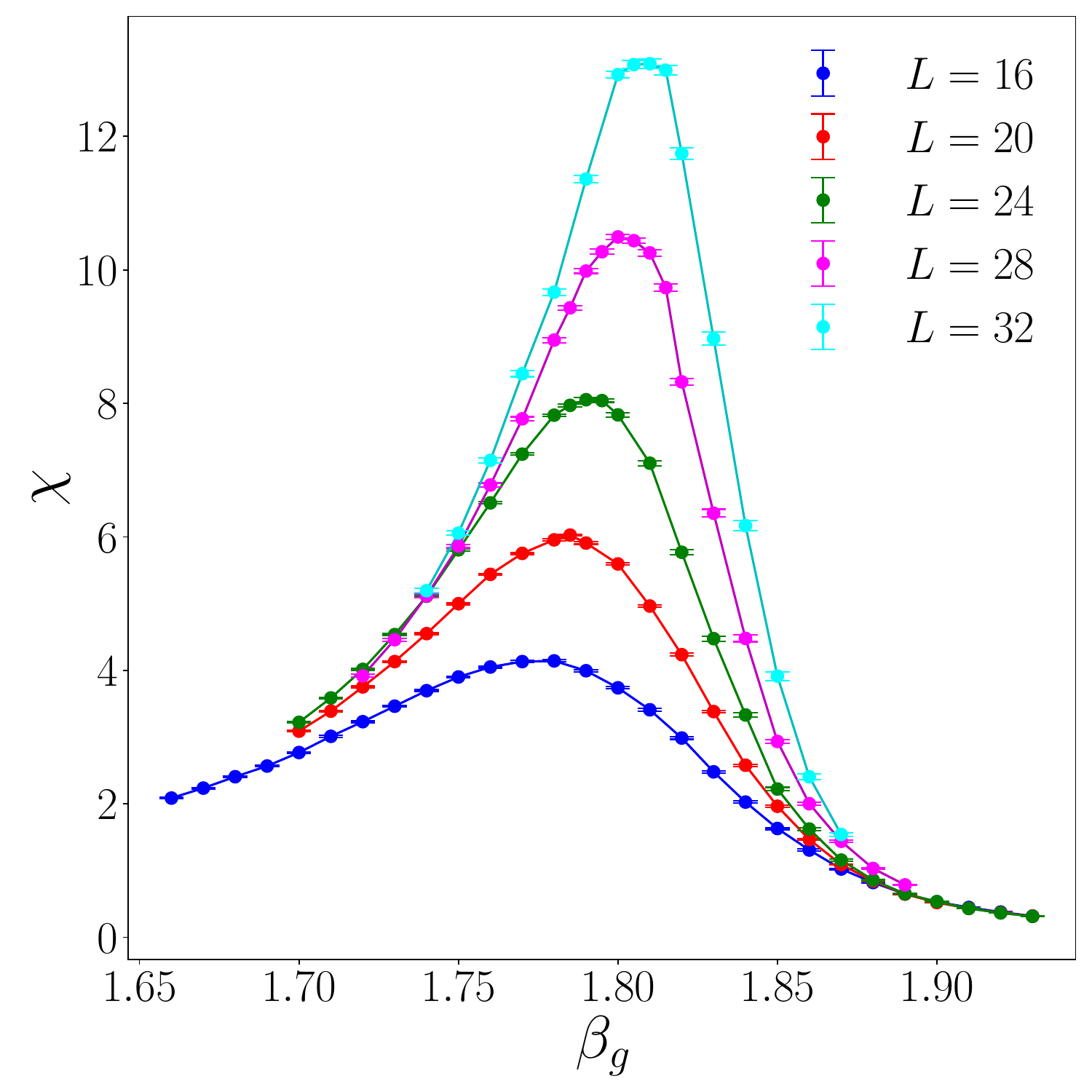}
  \caption{Susceptibility $\chi$ of the magnetic flux $\Gamma$ as a function of the coupling constant $\beta_g$ for $\beta_H=2.0$, and $V=L^4=16^4,20^4,24^4,28^4$ and $32^4$. The susceptibility increases as the lattice size $L$ increases, which is a sign of the phase transition.
  }
  \label{fig:chi}
\end{figure}

To quantify the properties of the phase transition such as its order, and if it is second order, to determine its universality class, we performed the finite-size scaling analysis. 
Following the scenario proposed by Motrunich and Senthil~\cite{motrunich2005origin}, we assume the Ising universality class for the finite-size scaling analysis.
To this end, we define the susceptibility $\chi$ from fluctuations of $\Gamma$ as
\bea
m &=& \frac{1}{N_3N_4}\sum_{x_3,x_4}\Gamma(x_3,x_4) , \\
\frac{\chi}{N_3N_4} &=&\langle m^2\rangle-\langle |m|\rangle^2 ,
\eea
identifying $\Gamma$ as the magnetization.
Similarly, we can define the susceptibility for the anti-vortex placed at $(x_1,x_2)=(N_1-{1}/{2},{(N_2-1)}/{2})$.
We emphasize here that the susceptibility $\chi$ is a two-dimensional one, while the lattice simulation is performed in four-dimensional spacetime.
We computed $\chi$ for $\beta_H=2.0$, and $V=L^4=16^4,20^4,24^4,28^4,32^4$ with changing $\beta_g$.
We computed $\chi$ for both vortex and anti-vortex, and averaged the results to reduce statistical fluctuations.

\begin{figure}[t]
  \centering
  \includegraphics[width=0.7\linewidth]{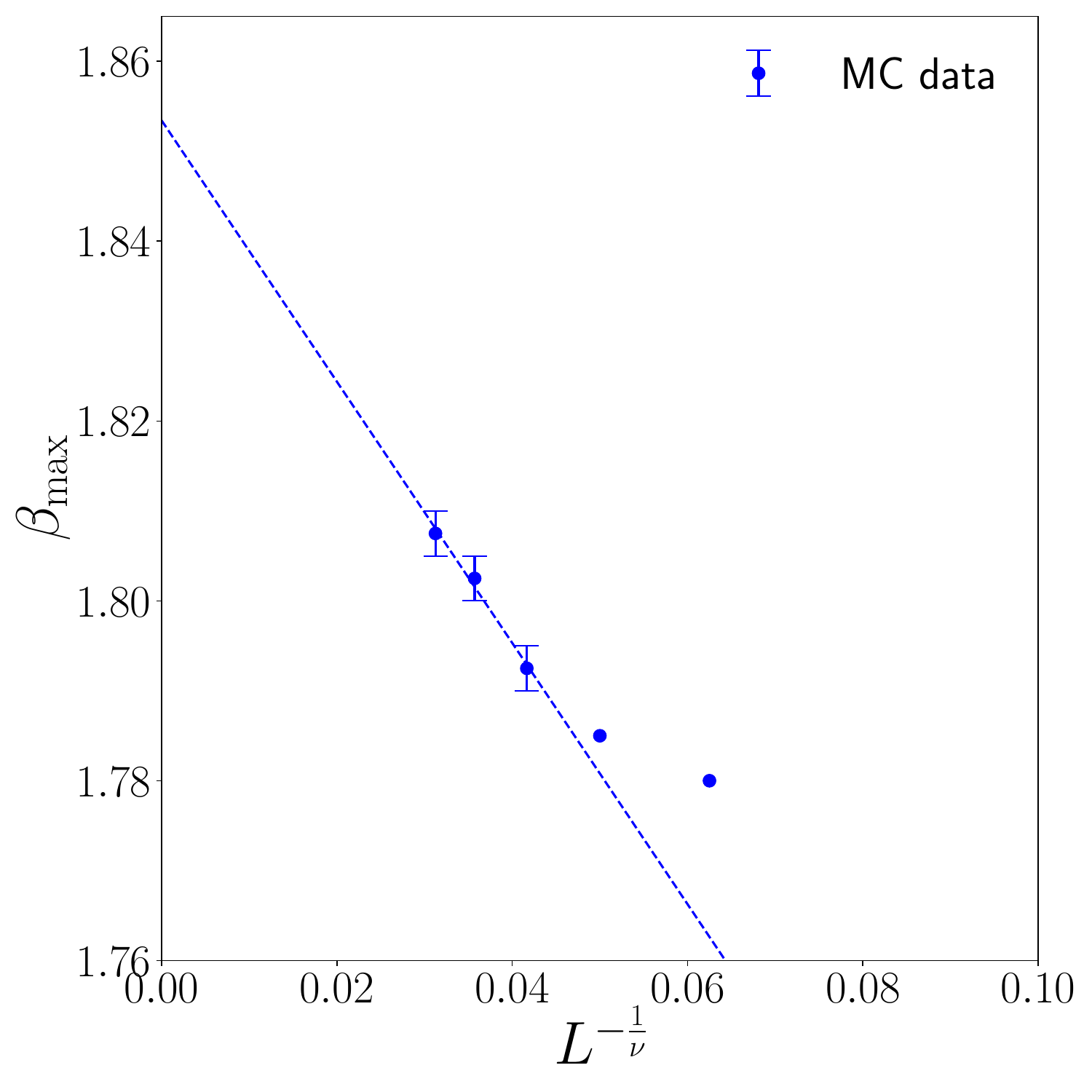}
  \caption{Finite-size scaling analysis for estimating the critical coupling constant in the infinite volume limit. The Monte Carlo results for $\beta_H=2.0$, and the system size $L=24$, $28$ and $32$ are fitted by $\beta_\mathrm{max}=\beta_c-aL^{-{1}/{\nu}}$, where $\beta_\mathrm{max}$ is the location of the maximum of $\chi$, and $\nu=1$ for the Ising universality class in two dimensions.
  }
  \label{fig:beta}
\end{figure}

\begin{figure}[t]
  \centering
  \includegraphics[width=0.7\linewidth]{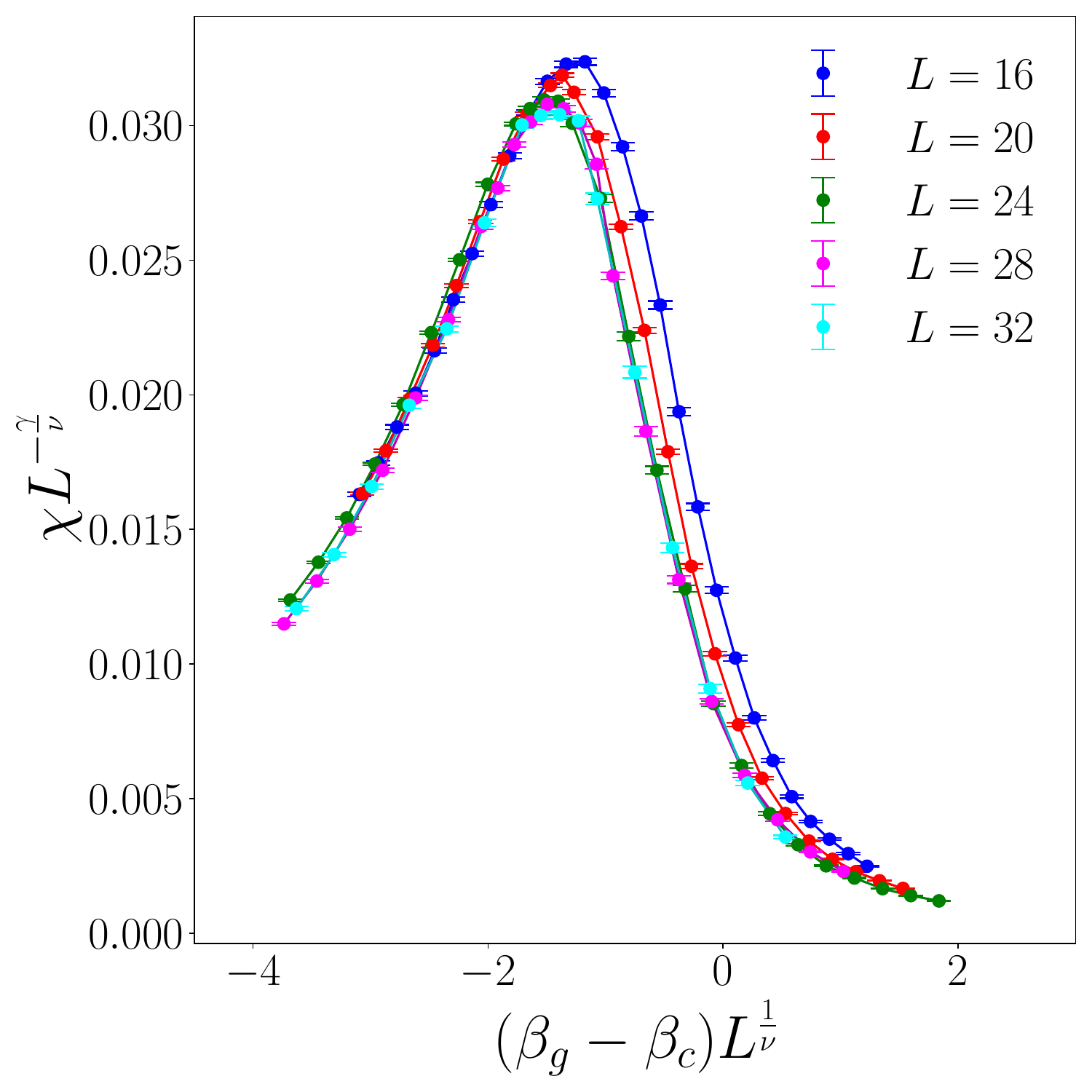}
  \caption{Renormalized susceptibility $\chi L^{-{\gamma}/{\nu}}$ as a function of the renormalized coupling constant $(\beta_g-\beta_c)^{{1}/{\nu}}$ for $\beta_H=2.0$, and $V=L^4=24^4$, $28^4$ and $32^4$ ($V=16^4$, $20^4$ are also shown just for information). We assume the Ising universality class in two dimensions, and used $\nu=1$ and $\gamma={7}/{4}$. All data collapse into the single curve implies that the observed phase transition belongs to the Ising universality class.
  }
  \label{fig:chi_renormalized}
\end{figure}

We show the susceptibility $\chi$ as a function of $\beta_g$ in figure~\ref{fig:chi}. We see that the maximum height of $\chi$ increases and the width of $\chi$ becomes narrower as $L$ increases, which would be a sign of the phase transition.
The results of the finite-size scaling are shown in figures~\ref{fig:beta} and~\ref{fig:chi_renormalized}.
First, we determine the critical coupling constant in the infinite volume limit, $\beta_c$ using the finite-size scaling. We fit the location of maximum values of $\chi$, $\beta_\mathrm{max}$ by $\beta_c-L^{-{1}/{\nu}}$ ($\nu=1$ for the two-dimensional Ising universality class). The fitting results are shown in figure~\ref{fig:beta}. We used $L=24,28,32$ for the fitting, and $\beta_c$ is estimated to be $\beta_c=1.853\pm0.006$.
Using this value and the critical exponents, we rescale $\beta_g$ and $\chi$. The renormalized dependence is shown in figure~\ref{fig:chi_renormalized}.
We plot $\chi L^{-{\gamma}/{\nu}}$ with $\gamma={7}/{4}$ as a function of $(\beta_g-\beta_c)^{{1}/{\nu}}$.
If the observed phase transition belongs to the Ising universality class, this renormalization will remove the volume dependence of $\chi$.
In fact, all data with different lattice sizes are collapsed into a single curve in figure~\ref{fig:chi_renormalized} strongly supports that the observed phase transition is actually second order, and its critical properties align with the Ising universality class as conjectured by Motrunich and Senthil. Finally, let us emphasize again that we confirm, for the first time, through an ab-initio simulation, exotic states of matter that can be distinguishable only by the phase transition occurring on extended objects.


\section{Summary}\label{sec:summary}

We have proposed a novel classification method to distinguish states of matter in the case that the bulk system changes smoothly without a phase transition based on spontaneous symmetry breaking in extended objects such as topological defects.
As an example, we have studied the phase transition on a vortex in a $\mathrm{U}(1)_{\mathrm{gauge}} \times \mathrm{U}(1)_{\mathrm{global}}$ lattice model in $(3+1)$ dimension, where the effective theory is a ($1+1$)-dimensional field theory with $\mathbb{Z}_2$ global symmetry. 
Even though a crossover is expected in terms of bulk observables, the two states may still be distinguishable through a vortex as a topological defect, and by observing a phase transition on it. In other words, the two states can be classified by spontaneous symmetry breaking of $\mathbb{Z}_2$ global symmetry on a votrtex. 
What kind of topological defects one can use as a probe depends on the symmetry-breaking pattern of the bulk state and can be classified using a conventional method. 
It is important to construct the low-energy effective theory of topological defect to discuss the possible state based on the phase transition in a topological defect.

We have considered a simple vortex; in general, in addition to extended objects like domain walls and vortices, we can consider junctions of domain walls, where a phase transition may also occur. It is also possible to consider line-like objects, such as Wilson loops, in which case the effective theory becomes a (0+1)-dimensional field theory, i.e., quantum mechanics. Although a strict phase transition does not occur in quantum mechanics, phenomena similar to level crossing may occur.

Regarding the hadron-quark continuity in finite-density QCD, our results are not directly applicable since the effective theory of CFL vortex is $\mathbb{C}P^2$ model according to the symmetry breaking pattern \cite{Nakano:2007dr,Eto:2009bh,Eto:2009tr,Eto:2021nle}, which has no discrete symmetry, and spontaneous symmetry breaking is not allowed by Coleman-Mermin-Wagner theorem \cite{PhysRev.158.383,PhysRevLett.17.1133}. 
On the other hand, gapless majorana fermions exist on a CFL vortex~\cite{Yasui:2010yw}, and thus the low-energy effective theory of CFL vortex may not be the simple $\mathbb{C}P^2$ model~\cite{Chatterjee:2016ykq}.
There is still a possibility that the interaction with fermions changes the phase structure of the $\mathbb{C}P^2$ model.
We will study it in future works.


\section*{Acknowledgements}
Y.H. thanks Yuya Tanizaki, Yui Hayashi, and Eiji Muto for their valuable discussions.
A part of the numerical simulations has been performed using cluster computers at iTHEMS in RIKEN, and Yukawa-21 at YITP in Kyoto University.
This work was supported by JSPS KAKENHI Grant Numbers~JP21H01084, JP24H00975, JP24K00630 and JP24KJ0613.


\appendix

\section{Ising model}\label{sec:Ising}
As a concrete example satisfying the criterion of symmetry breaking in eq.~\eqref{eq:criteria}, let us consider an $N$-site quantum Ising model with a transverse magnetic field in $(1+1)$ dimensions, whose Hamiltonian  is given by
\begin{align}
{H}&=-J\sum_{i=0}^{N-1} \sigma^z_{i} \sigma^z_{i+1}-h\sum_{i=0}^{N-1} \sigma^x_{i}.
\end{align}
Here, $J$ and $h$ are the coupling constant and the transverse magnetic field, respectively (we assume $J>0$ and $h>0$ for simplicity). $\sigma_{i}^z$ and $\sigma_{i}^x$ are the Pauli $z$ and $x$ operators, respectively.
We assume the periodic boundary condition, and the $N$th site is identified with the $0$th site. This model has a $\mathbb{Z}_2$ symmetry that flips spins $\sigma_{i}^z \rightarrow -\sigma_{i}^z$ for all $i$ simultaneously.
This model exhibits an ordered phase ($J \gg h$) characterized by spontaneous $\mathbb{Z}_2$ symmetry breaking and long-range spin alignment, and a disordered phase ($h \gg J$) where the symmetry is restored and spins align along the transverse field, but no long-range order is present~\cite{Sachdev:2011fcc}.

In order to detect the broken phase, we introduce a symmetry defect between the sites $N-1$ and $N$ ($=0$), which flips the coupling constant of $\sigma_{N-1}\sigma_0$, 
\begin{equation}
  \begin{split}
    {H}'&=+J \sigma^z_{N-1} \sigma^z_{0}
    -J\sum_{i=0}^{N-2} \sigma^z_{i} \sigma^z_{i+1}
    -h\sum_{i=0}^{N-1} \sigma^x_{i}\\
    &= 2J\sigma^z_{N-1} \sigma^z_{0} +H.
  \end{split}
\end{equation}
We now explore the ground state solutions for strong $J\gg h$ and weak $J\ll h$ couplings.
In the strong coupling limit where we can take $h=0$, the ground states are doubly degenerate, and given explicitly as $\ket{\Omega_{\uparrow}}=\prod_i\ket{\uparrow}_i$ and $\ket{\Omega_{\downarrow}}=\prod_i\ket{\downarrow}_i$, in the absence of the symmetry defect.
Their energies are given by
\begin{equation}
H \ket{\Omega_{\uparrow}} = -JN \ket{\Omega_{\uparrow}}, \quad
H \ket{\Omega_{\downarrow}} = -JN \ket{\Omega_{\downarrow}}.
\end{equation}
In this case, the partition function behaves as
\begin{equation}
Z=\tr e^{-\beta H}\simeq 
\bra{\Omega_{\uparrow}}e^{-\beta H}\ket{\Omega_{\uparrow}}
+\bra{\Omega_{\downarrow}}e^{-\beta H}\ket{\Omega_{\downarrow}}
=2e^{\beta JN}
\end{equation}
for large $\beta$.

On the other hand, ground states of ${H}'$ are
\begin{align}
\ket{\Omega'_{\uparrow k}} &= \ket{\uparrow}_{0}\ket{\uparrow}_{1}\ket{\uparrow}_{2}\cdots \ket{\uparrow}_{k}\ket{\downarrow}_{k+1}\cdots \ket{\downarrow}_{N-1},\\
\ket{\Omega'_{\downarrow k}} &= \ket{\downarrow}_{0}\ket{\downarrow}_{1}\ket{\downarrow}_{2}\cdots \ket{\downarrow}_{k}\ket{\uparrow}_{k+1}\cdots \ket{\uparrow}_{N-1},
\end{align}
where a domain wall located between the sites $k$ and $k+1$; $k$ runs from $0$ to $N-1$.
The corresponding energies are
\begin{align}
H' \ket{\Omega_{\uparrow k }} &= -J(N-2) \ket{\Omega_{\uparrow k }}, \\
H' \ket{\Omega_{\downarrow k }} &= -J(N-2) \ket{\Omega_{\downarrow k }}.
\end{align}
Therefore, the partition function in the existence of symmetry defect reads
\begin{equation}
Z'=    \tr e^{-\beta H'}\simeq
\sum_{k=0}^{N-1}\qty( \bra{\Omega'_{\uparrow k}} e^{-\beta {H}'}\ket{\Omega'_{\uparrow k}}
+\bra{\Omega'_{\downarrow k}} e^{-\beta {H}'}\ket{\Omega'_{\downarrow k}}
)
=
2N e^{\beta J(N-2)}.
\end{equation}
By taking the ratio, we obtain
\begin{equation}
\lim_{\beta\to\infty}\frac{Z'}{Z}= N e^{-2\beta J }\to 0,
\end{equation}
which is the signal of spontaneous breaking of $\mathbb{Z}_2$ symmetry.

Next, consider a weak coupling case, $h\gg J$, where we neglect $J$.
Since the interaction between spins is negligible, the ground state wavefunction can be written in terms of the product of the eigenstates of $\sigma^x$,
\begin{equation}
\ket{\Omega} = \prod_{i=0}^{N-1}\ket{+}_i,
\end{equation}
where we define
\begin{equation}
\ket{+}\coloneqq\frac{1}{\sqrt{2}}  (\ket{\uparrow} + \ket{\downarrow} ),
\end{equation}
which is the eigenstate of $\sigma^x$ operator, $\sigma^x\ket{+}=+\ket{+}$.
This state is an eigenstate of both Hamiltonians with and without the symmetry defect in the weak coupling limit,
\begin{align}
    H\ket{\Omega}=-hN \ket{\Omega},\quad
     H'\ket{\Omega}=-hN \ket{\Omega},
\end{align}
which leads to
\begin{align}
    \bra{\Omega}\mathrm{e}^{-\beta H}\ket{\Omega}=\mathrm{e}^{\beta hN},\quad
    \bra{\Omega}\mathrm{e}^{-\beta H'}\ket{\Omega}=\mathrm{e}^{\beta hN}.
\end{align}
Therefore, we obtain
\begin{equation}
\lim_{\beta\rightarrow\infty}\frac{Z'}{Z}=
    \lim_{\beta\rightarrow\infty}\frac{\operatorname{tr}\mathrm{e}^{-\beta H'}}{\operatorname{tr} \mathrm{e}^{-\beta H}}
    =1,
\end{equation}
which is the signal of unbroken phase of $\mathbb{Z}_2$ symmetry.
By introducing the symmetry defect, we confirmed that the criteria \eqref{eq:criteria} works.

\bibliographystyle{JHEP}
\bibliography{main}

\providecommand{\href}[2]{#2}\begingroup\raggedright\begin{thebibliography}{10}

\bibitem{Fukushima:2010bq}
K.~Fukushima and T.~Hatsuda, \emph{{The phase diagram of dense QCD}},
  \href{https://doi.org/10.1088/0034-4885/74/1/014001}{\emph{Rept. Prog. Phys.}
  {\bfseries 74} (2011) 014001}
  [\href{https://arxiv.org/abs/1005.4814}{{\ttfamily 1005.4814}}].

\bibitem{Baym:2017whm}
G.~Baym, T.~Hatsuda, T.~Kojo, P.D.~Powell, Y.~Song and T.~Takatsuka,
  \emph{{From hadrons to quarks in neutron stars: a review}},
  \href{https://doi.org/10.1088/1361-6633/aaae14}{\emph{Rept. Prog. Phys.}
  {\bfseries 81} (2018) 056902}
  [\href{https://arxiv.org/abs/1707.04966}{{\ttfamily 1707.04966}}].

\bibitem{Landau:1937obd}
L.D.~Landau, \emph{{On the theory of phase transitions}},
  \href{https://doi.org/10.1016/B978-0-08-010586-4.50034-1}{\emph{Zh. Eksp.
  Teor. Fiz.} {\bfseries 7} (1937) 19}.

\bibitem{Ginzburg:1950sr}
V.L.~Ginzburg and L.D.~Landau, \emph{{On the Theory of superconductivity}},
  \href{https://doi.org/10.1016/b978-0-08-010586-4.50078-x}{\emph{Zh. Eksp.
  Teor. Fiz.} {\bfseries 20} (1950) 1064}.

\bibitem{Nambu:1960tm}
Y.~Nambu, \emph{{Quasiparticles and Gauge Invariance in the Theory of
  Superconductivity}},
  \href{https://doi.org/10.1103/PhysRev.117.648}{\emph{Phys. Rev.} {\bfseries
  117} (1960) 648}.

\bibitem{Nambu:1961fr}
Y.~Nambu and G.~Jona-Lasinio, \emph{{Dynamical model of elementary particles
  based on an analogy with superconductivity. II.}},
  \href{https://doi.org/10.1103/PhysRev.124.246}{\emph{Phys. Rev.} {\bfseries
  124} (1961) 246}.

\bibitem{Alford:1997zt}
M.G.~Alford, K.~Rajagopal and F.~Wilczek, \emph{{QCD at finite baryon density:
  Nucleon droplets and color superconductivity}},
  \href{https://doi.org/10.1016/S0370-2693(98)00051-3}{\emph{Phys. Lett. B}
  {\bfseries 422} (1998) 247}
  [\href{https://arxiv.org/abs/hep-ph/9711395}{{\ttfamily hep-ph/9711395}}].

\bibitem{Alford:1998mk}
M.G.~Alford, K.~Rajagopal and F.~Wilczek, \emph{{Color flavor locking and
  chiral symmetry breaking in high density QCD}},
  \href{https://doi.org/10.1016/S0550-3213(98)00668-3}{\emph{Nucl. Phys. B}
  {\bfseries 537} (1999) 443}
  [\href{https://arxiv.org/abs/hep-ph/9804403}{{\ttfamily hep-ph/9804403}}].

\bibitem{Alford:2007xm}
M.G.~Alford, A.~Schmitt, K.~Rajagopal and T.~Sch\"afer, \emph{{Color
  superconductivity in dense quark matter}},
  \href{https://doi.org/10.1103/RevModPhys.80.1455}{\emph{Rev. Mod. Phys.}
  {\bfseries 80} (2008) 1455}
  [\href{https://arxiv.org/abs/0709.4635}{{\ttfamily 0709.4635}}].

\bibitem{Schafer:1998ef}
T.~Sch\"afer and F.~Wilczek, \emph{{Continuity of quark and hadron matter}},
  \href{https://doi.org/10.1103/PhysRevLett.82.3956}{\emph{Phys. Rev. Lett.}
  {\bfseries 82} (1999) 3956}
  [\href{https://arxiv.org/abs/hep-ph/9811473}{{\ttfamily hep-ph/9811473}}].

\bibitem{Schafer:1999fe}
T.~Sch\"afer, \emph{{Patterns of symmetry breaking in QCD at high baryon
  density}}, \href{https://doi.org/10.1016/S0550-3213(00)00063-8}{\emph{Nucl.
  Phys. B} {\bfseries 575} (2000) 269}
  [\href{https://arxiv.org/abs/hep-ph/9909574}{{\ttfamily hep-ph/9909574}}].

\bibitem{Fradkin:1978dv}
E.H.~Fradkin and S.H.~Shenker, \emph{{Phase Diagrams of Lattice Gauge Theories
  with Higgs Fields}},
  \href{https://doi.org/10.1103/PhysRevD.19.3682}{\emph{Phys. Rev. D}
  {\bfseries 19} (1979) 3682}.

\bibitem{Banks:1979fi}
T.~Banks and E.~Rabinovici, \emph{{Finite Temperature Behavior of the Lattice
  Abelian Higgs Model}},
  \href{https://doi.org/10.1016/0550-3213(79)90064-6}{\emph{Nucl. Phys. B}
  {\bfseries 160} (1979) 349}.

\bibitem{OSTERWALDER1978440}
K.~Osterwalder and E.~Seiler, \emph{Gauge field theories on a lattice},
  \href{https://doi.org/https://doi.org/10.1016/0003-4916(78)90039-8}{\emph{Annals
  of Physics} {\bfseries 110} (1978) 440}.

\bibitem{Cherman:2018jir}
A.~Cherman, S.~Sen and L.G.~Yaffe, \emph{{Anyonic particle-vortex statistics
  and the nature of dense quark matter}},
  \href{https://doi.org/10.1103/PhysRevD.100.034015}{\emph{Phys. Rev. D}
  {\bfseries 100} (2019) 034015}
  [\href{https://arxiv.org/abs/1808.04827}{{\ttfamily 1808.04827}}].

\bibitem{Hirono:2018fjr}
Y.~Hirono and Y.~Tanizaki, \emph{{Quark-Hadron Continuity beyond the
  Ginzburg-Landau Paradigm}},
  \href{https://doi.org/10.1103/PhysRevLett.122.212001}{\emph{Phys. Rev. Lett.}
  {\bfseries 122} (2019) 212001}
  [\href{https://arxiv.org/abs/1811.10608}{{\ttfamily 1811.10608}}].

\bibitem{Hirono:2019oup}
Y.~Hirono and Y.~Tanizaki, \emph{{Effective gauge theories of superfluidity
  with topological order}},
  \href{https://doi.org/10.1007/JHEP07(2019)062}{\emph{JHEP} {\bfseries 07}
  (2019) 062} [\href{https://arxiv.org/abs/1904.08570}{{\ttfamily
  1904.08570}}].

\bibitem{Cherman:2020hbe}
A.~Cherman, T.~Jacobson, S.~Sen and L.G.~Yaffe, \emph{{Higgs-confinement phase
  transitions with fundamental representation matter}},
  \href{https://doi.org/10.1103/PhysRevD.102.105021}{\emph{Phys. Rev. D}
  {\bfseries 102} (2020) 105021}
  [\href{https://arxiv.org/abs/2007.08539}{{\ttfamily 2007.08539}}].

\bibitem{Hidaka:2022blq}
Y.~Hidaka and D.~Kondo, \emph{{Emergent higher-form symmetry in Higgs phases
  with superfluidity}},  \href{https://arxiv.org/abs/2210.11492}{{\ttfamily
  2210.11492}}.

\bibitem{Verresen:2022mcr}
R.~Verresen, U.~Borla, A.~Vishwanath, S.~Moroz and R.~Thorngren, \emph{{Higgs
  Condensates are Symmetry-Protected Topological Phases: I. Discrete
  Symmetries}},  \href{https://arxiv.org/abs/2211.01376}{{\ttfamily
  2211.01376}}.

\bibitem{Thorngren:2023ple}
R.~Thorngren, T.~Rakovszky, R.~Verresen and A.~Vishwanath, \emph{{Higgs
  Condensates are Symmetry-Protected Topological Phases: II. $U(1)$ Gauge
  Theory and Superconductors}},
  \href{https://arxiv.org/abs/2303.08136}{{\ttfamily 2303.08136}}.

\bibitem{Chung:2024hsq}
K.T.K.~Chung, R.~Flores-Calder\'on, R.C.~Torres, P.~Ribeiro, S.~Moroz and
  P.~McClarty, \emph{{Higgs Phases and Boundary Criticality}},
  \href{https://arxiv.org/abs/2404.17001}{{\ttfamily 2404.17001}}.

\bibitem{Hayashi:2023sas}
Y.~Hayashi, \emph{{Higgs-Confinement Continuity and Matching of Aharonov-Bohm
  Phases}}, \href{https://doi.org/10.1103/PhysRevLett.132.221901}{\emph{Phys.
  Rev. Lett.} {\bfseries 132} (2024) 221901}
  [\href{https://arxiv.org/abs/2303.02129}{{\ttfamily 2303.02129}}].

\bibitem{Cherman:2024exo}
A.~Cherman, T.~Jacobson, S.~Sen and L.G.~Yaffe, \emph{{Line operators, vortex
  statistics, and Higgs versus confinement dynamics}},
  \href{https://doi.org/10.1007/JHEP06(2024)200}{\emph{JHEP} {\bfseries 06}
  (2024) 200} [\href{https://arxiv.org/abs/2401.17489}{{\ttfamily
  2401.17489}}].

\bibitem{Alford:2018mqj}
M.G.~Alford, G.~Baym, K.~Fukushima, T.~Hatsuda and M.~Tachibana,
  \emph{{Continuity of vortices from the hadronic to the color-flavor locked
  phase in dense matter}},
  \href{https://doi.org/10.1103/PhysRevD.99.036004}{\emph{Phys. Rev. D}
  {\bfseries 99} (2019) 036004}
  [\href{https://arxiv.org/abs/1803.05115}{{\ttfamily 1803.05115}}].

\bibitem{Chatterjee:2018nxe}
C.~Chatterjee, M.~Nitta and S.~Yasui, \emph{{Quark-hadron continuity under
  rotation: Vortex continuity or boojum?}},
  \href{https://doi.org/10.1103/PhysRevD.99.034001}{\emph{Phys. Rev. D}
  {\bfseries 99} (2019) 034001}
  [\href{https://arxiv.org/abs/1806.09291}{{\ttfamily 1806.09291}}].

\bibitem{Fujimoto:2019sxg}
Y.~Fujimoto, K.~Fukushima and W.~Weise, \emph{{Continuity from neutron matter
  to two-flavor quark matter with $^1 S_0$ and $^3 P_2$ superfluidity}},
  \href{https://doi.org/10.1103/PhysRevD.101.094009}{\emph{Phys. Rev. D}
  {\bfseries 101} (2020) 094009}
  [\href{https://arxiv.org/abs/1908.09360}{{\ttfamily 1908.09360}}].

\bibitem{Fujimoto:2020dsa}
Y.~Fujimoto and M.~Nitta, \emph{{Non-Abelian Alice strings in two-flavor dense
  QCD}}, \href{https://doi.org/10.1103/PhysRevD.103.054002}{\emph{Phys. Rev. D}
  {\bfseries 103} (2021) 054002}
  [\href{https://arxiv.org/abs/2011.09947}{{\ttfamily 2011.09947}}].

\bibitem{Fujimoto:2021wsr}
Y.~Fujimoto and M.~Nitta, \emph{{Topological confinement of vortices in
  two-flavor dense QCD}},
  \href{https://doi.org/10.1007/JHEP09(2021)192}{\emph{JHEP} {\bfseries 09}
  (2021) 192} [\href{https://arxiv.org/abs/2103.15185}{{\ttfamily
  2103.15185}}].

\bibitem{Fujimoto:2021bes}
Y.~Fujimoto and M.~Nitta, \emph{{Vortices penetrating two-flavor quark-hadron
  continuity}}, \href{https://doi.org/10.1103/PhysRevD.103.114003}{\emph{Phys.
  Rev. D} {\bfseries 103} (2021) 114003}
  [\href{https://arxiv.org/abs/2102.12928}{{\ttfamily 2102.12928}}].

\bibitem{Greensite:2017ajx}
J.~Greensite and K.~Matsuyama, \emph{{Confinement criterion for gauge theories
  with matter fields}},
  \href{https://doi.org/10.1103/PhysRevD.96.094510}{\emph{Phys. Rev. D}
  {\bfseries 96} (2017) 094510}
  [\href{https://arxiv.org/abs/1708.08979}{{\ttfamily 1708.08979}}].

\bibitem{Greensite:2018mhh}
J.~Greensite and K.~Matsuyama, \emph{{What symmetry is actually broken in the
  Higgs phase of a gauge-Higgs theory?}},
  \href{https://doi.org/10.1103/PhysRevD.98.074504}{\emph{Phys. Rev. D}
  {\bfseries 98} (2018) 074504}
  [\href{https://arxiv.org/abs/1805.00985}{{\ttfamily 1805.00985}}].

\bibitem{Greensite:2020nhg}
J.~Greensite and K.~Matsuyama, \emph{{Higgs phase as a spin glass and the
  transition between varieties of confinement}},
  \href{https://doi.org/10.1103/PhysRevD.101.054508}{\emph{Phys. Rev. D}
  {\bfseries 101} (2020) 054508}
  [\href{https://arxiv.org/abs/2001.03068}{{\ttfamily 2001.03068}}].

\bibitem{Greensite:2021fyi}
J.~Greensite and K.~Matsuyama, \emph{{Symmetry, Confinement, and the Higgs
  Phase}}, \href{https://doi.org/10.3390/sym14010177}{\emph{Symmetry}
  {\bfseries 14} (2022) 177}
  [\href{https://arxiv.org/abs/2112.06421}{{\ttfamily 2112.06421}}].

\bibitem{Ikeda:2023kcf}
R.~Ikeda, S.~Kato, K.-I.~Kondo and A.~Shibata, \emph{{Gauge-independent
  transition separating confinement and Higgs phases~in lattice SU(2) gauge
  theory with a scalar field in~the~fundamental~representation}},
  \href{https://doi.org/10.1103/PhysRevD.109.054505}{\emph{Phys. Rev. D}
  {\bfseries 109} (2024) 054505}
  [\href{https://arxiv.org/abs/2308.13430}{{\ttfamily 2308.13430}}].

\bibitem{Balachandran:2005ev}
A.P.~Balachandran, S.~Digal and T.~Matsuura, \emph{{Semi-superfluid strings in
  high density QCD}},
  \href{https://doi.org/10.1103/PhysRevD.73.074009}{\emph{Phys. Rev. D}
  {\bfseries 73} (2006) 074009}
  [\href{https://arxiv.org/abs/hep-ph/0509276}{{\ttfamily hep-ph/0509276}}].

\bibitem{Nakano:2008dc}
E.~Nakano, M.~Nitta and T.~Matsuura, \emph{{Non-Abelian Strings in Hot or Dense
  QCD}}, \href{https://doi.org/10.1143/PTPS.174.254}{\emph{Prog. Theor. Phys.
  Suppl.} {\bfseries 174} (2008) 254}
  [\href{https://arxiv.org/abs/0805.4539}{{\ttfamily 0805.4539}}].

\bibitem{Eto:2009kg}
M.~Eto and M.~Nitta, \emph{{Color Magnetic Flux Tubes in Dense QCD}},
  \href{https://doi.org/10.1103/PhysRevD.80.125007}{\emph{Phys. Rev. D}
  {\bfseries 80} (2009) 125007}
  [\href{https://arxiv.org/abs/0907.1278}{{\ttfamily 0907.1278}}].

\bibitem{Eto:2013hoa}
M.~Eto, Y.~Hirono, M.~Nitta and S.~Yasui, \emph{{Vortices and Other Topological
  Solitons in Dense Quark Matter}},
  \href{https://doi.org/10.1093/ptep/ptt095}{\emph{PTEP} {\bfseries 2014}
  (2014) 012D01} [\href{https://arxiv.org/abs/1308.1535}{{\ttfamily
  1308.1535}}].

\bibitem{Nakano:2007dr}
E.~Nakano, M.~Nitta and T.~Matsuura, \emph{{Non-Abelian strings in high density
  QCD: Zero modes and interactions}},
  \href{https://doi.org/10.1103/PhysRevD.78.045002}{\emph{Phys. Rev. D}
  {\bfseries 78} (2008) 045002}
  [\href{https://arxiv.org/abs/0708.4096}{{\ttfamily 0708.4096}}].

\bibitem{motrunich2005origin}
O.~Motrunich and T.~Senthil, \emph{Origin of artificial electrodynamics in
  three-dimensional bosonic models}, {\emph{Physical Review B—Condensed
  Matter and Materials Physics} {\bfseries 71} (2005) 125102}.

\bibitem{Weinberg:1996kr}
S.~Weinberg, \emph{{The quantum theory of fields. Vol. 2: Modern
  applications}}, Cambridge University Press (8, 2013),
  \href{https://doi.org/10.1017/CBO9781139644174}{10.1017/CBO9781139644174}.

\bibitem{Eto:2009bh}
M.~Eto, E.~Nakano and M.~Nitta, \emph{{Effective world-sheet theory of color
  magnetic flux tubes in dense QCD}},
  \href{https://doi.org/10.1103/PhysRevD.80.125011}{\emph{Phys. Rev. D}
  {\bfseries 80} (2009) 125011}
  [\href{https://arxiv.org/abs/0908.4470}{{\ttfamily 0908.4470}}].

\bibitem{Eto:2009tr}
M.~Eto, M.~Nitta and N.~Yamamoto, \emph{{Instabilities of Non-Abelian Vortices
  in Dense QCD}},
  \href{https://doi.org/10.1103/PhysRevLett.104.161601}{\emph{Phys. Rev. Lett.}
  {\bfseries 104} (2010) 161601}
  [\href{https://arxiv.org/abs/0912.1352}{{\ttfamily 0912.1352}}].

\bibitem{Eto:2021nle}
M.~Eto and M.~Nitta, \emph{{Chiral non-Abelian vortices and their confinement
  in three flavor dense QCD}},
  \href{https://doi.org/10.1103/PhysRevD.104.094052}{\emph{Phys. Rev. D}
  {\bfseries 104} (2021) 094052}
  [\href{https://arxiv.org/abs/2103.13011}{{\ttfamily 2103.13011}}].

\bibitem{PhysRev.158.383}
P.C.~Hohenberg, \emph{Existence of long-range order in one and two dimensions},
  \href{https://doi.org/10.1103/PhysRev.158.383}{\emph{Phys. Rev.} {\bfseries
  158} (1967) 383}.

\bibitem{PhysRevLett.17.1133}
N.D.~Mermin and H.~Wagner, \emph{Absence of ferromagnetism or
  antiferromagnetism in one- or two-dimensional isotropic heisenberg models},
  \href{https://doi.org/10.1103/PhysRevLett.17.1133}{\emph{Phys. Rev. Lett.}
  {\bfseries 17} (1966) 1133}.

\bibitem{Yasui:2010yw}
S.~Yasui, K.~Itakura and M.~Nitta, \emph{{Fermion structure of non-Abelian
  vortices in high density QCD}},
  \href{https://doi.org/10.1103/PhysRevD.81.105003}{\emph{Phys. Rev. D}
  {\bfseries 81} (2010) 105003}
  [\href{https://arxiv.org/abs/1001.3730}{{\ttfamily 1001.3730}}].

\bibitem{Chatterjee:2016ykq}
C.~Chatterjee, M.~Cipriani and M.~Nitta, \emph{{Coupling between Majorana
  fermions and Nambu-Goldstone bosons inside a non-Abelian vortex in dense
  QCD}}, \href{https://doi.org/10.1103/PhysRevD.93.065046}{\emph{Phys. Rev. D}
  {\bfseries 93} (2016) 065046}
  [\href{https://arxiv.org/abs/1602.01677}{{\ttfamily 1602.01677}}].

\bibitem{Sachdev:2011fcc}
S.~Sachdev, \emph{{Quantum Phase Transitions}}, Cambridge University Press (4,
  2011),
  \href{https://doi.org/10.1017/cbo9780511973765}{10.1017/cbo9780511973765}.

\end{thebibliography}\endgroup

\end{document}